\documentclass[11pt]{article}
\usepackage{mathrsfs}
\usepackage{amsfonts}
\usepackage{amsmath,amsthm,hyperref}
\usepackage{amssymb}
\usepackage{hyperref,color}
\usepackage{exscale,relsize,graphics}
\newtheorem{thm}{Theorem}[subsection]

\newtheorem{prop}[thm]{Proposition}
\theoremstyle{definition}

\theoremstyle{remark}
\newtheorem{rem}[thm]{Remark}
\numberwithin{equation}{subsection}

\newcommand{\eql}[2]{\begin{equation}\label{#1}#2\end{equation}}

\newcommand{\eqnarrayn}[1]{\begin{eqnarray*}#1\end{eqnarray*}}

\def\cur{\mathcal{K}_n}

\begin{document}
\author{ Peng Zhao \and Engui Fan\footnote{Corresponding
author and  e-mail address:
      faneg@fudan.edu.cn}}
\date{   \small{ School of Mathematical Sciences, Institute of Mathematics \\
 and Key Laboratory of Mathematics for Nonlinear Science, \\ Fudan
University, Shanghai 200433, P.R. China}}
\title{\bf \Large{Reality problems for the Algebro-Geometric Solutions of Fokas-Lenell hierarchy} }
\maketitle

\begin{abstract}
  In a previous study, we obtained the algebro-geometric solutions and
  $n$-dark solitons
  of Forkas-Lenells (FL) hierarchy
  using algebro-geometric method.
  In this paper, we
  construct  physically relevant classes of solutions for FL hierarchy 
  by studying the reality conditions for $q=\pm \bar{r}$ based on  the idea of Vinikov's homological basis. 
\end{abstract}
\section*{Introduction}

The Forkas-Lenells (FL) system \cite{For,Len,Len2,Len3}
\begin{equation}\label{1.1}
\begin{split}
 &q_{xt}-q_{xx}+iqq_xr-2i q_x+q=0,\\
&r_{xt}-r_{xx}-iqrr_x+2i r_x+r=0,
\end{split}
\end{equation}
where $q,r$ is a complex-valued function of $x$ and $t$,
has been studied extensively in
relation with various aspects such as Lax integrability, bi-Hamiltonian
structure, various kinds of exact solutions, etc.\cite{VE,OC,YM,YM1,JS,05}.
In ref.\,\cite{05}, we have constructed the FL hierarchy  using polynomial recursion formalism,
then derived the algebro-geometric solutions of the whole hierarchy
and degenerated them into
$n$-dark solitons.
 To find solutions we go through an intermediate step,
  a complexified version of FL hierarchy.
  However,
  in physical applications,
  two natural reductions of (\ref{1.1}),
  the defocusing case, with $r=-\overline{q},$
  \begin{flalign}\label{1.1a}
   \textrm{FL}_-:\quad q_{xt}-q_{xx}-i|q|^2q_x-2i q_x+q=0,
  \end{flalign}
  and focusing case, with $r=\overline{q}$,
  \begin{flalign}\label{1.1b}
   \textrm{FL}_+:\quad q_{xt}-q_{xx}+i|q|^2q_x-2i q_x+q=0,
  \end{flalign}
  are of interest.
  Therefore, a natural question arising here
  is how to derive the solutions of FL
  from the solutions of (\ref{1.1}).
  From the modern mathematical point of view, the algebro-geometric solutions of FL$_{\pm}$ hierarchy
  can be more or less solved if the formulas
  (\ref{1227}) and (\ref{1228}) are derived.
  However, this correspondence is only formal to some extent and needs
  further
  analysis about the parameters in the solutions (\ref{1227}) and (\ref{1228}).
  As remark in \cite{04}, the one-soliton of AKNS system, depends on four complex
  parameters $a,b,\phi,\psi.$ Setting
\begin{flalign}
 \ell(x,t)=\exp\left(ax+f_0(a)t+\phi\right),\quad m(x,t)=\exp\left(bx-f_0(-b)t+\psi\right),
\end{flalign}
the one-soliton is given by
\begin{flalign}
 q=&(a+b)m/(1-\ell m),\\
 r=&(a+b)\ell/(1-\ell m).
\end{flalign}
It can easily be arranged at the denominator does not vanish, say, for $t=0$
and all $x\in\mathbb{R},$ but that there are singularities after a finite time $t_0.$
Hence the AKNS system is not specific enough to prevent solutions from exploding.
However, after the reduction $r=-\overline{q},$ we arrive at the NLS equation
and the one-soliton of NLS is
\begin{flalign}\label{a1.6}
 q(x,t)=-Re(a)e^{-Im(a)x+[Im(a)^2-Re(a)^2]t}cosh^{-1}(Re(a)[x+2Im(a)t]).
\end{flalign}
Solution (\ref{a1.6}) is smooth without any singularity.
The reason
why the solution of AKNS system shows different properties with that of
NLS equation
is obvious: the solutions of
AKNS system contains more parameters
and hence can be more complicated than those of NLS equation. Therefore,
finding useful formulas about parameters in solutions of FL hierarchy is an efficient
and important way to
obtain the solution of
FL$_{\pm}$ hierarchy.

The application of this idea to the algebro-geometric method
can be found in the wonderful works \cite{Its}. In \cite{05}
we have constructed the algebro-geometric solutions of FL hierarchy
using the algebro-geometric method \cite{15}.
The basic strategy we shall take
is to
apply the idea
of
\cite{Its}
to \cite{05} and
the
technique
developed
in this text
seems
be available
to solve the reality problems of
the algebro-geometric solutions
of
other integrable equations
via the algebro-geometric method \cite{15}.
Here we should emphasize
the work of the authors in \cite{15}.
They
have discussed the reduction conditions for the
algebro-geometric solutions of
AKNS hierarchy using Vinikov's basis (for the definition, see Proposition 1.1.1).
However, it may
be very complicated to depict the Vinikov's basis for a general real curve,
although this problem become simple in some special cases (e.g. all the
branch points of the spectral curve are real). Moreover,
various of path integration in explicit
algebro-geometric solutions
and the action of holomorphic involution on spectral curve
very often involves the specific path,
which inevitably connects with
the choice of homology basis. Thus, we shall
solve the reality problem by using very specific
curve and homology basis
and every parameter
in the finally
expressions
of solutions is very clear.
This is also
the biggest difference
from 'abstract' Vinikov's basis.

This paper is organized as follows. To be self-contained,
we give a fundamental discription of real Riemann surface theory and general
results about the algebro-geometric solutions of FL hierarchy
in section 1. In section 2 we obtain
two complementary results, which indicates the information of
the algebro-geometric solutions is totally
included in the spectal
data. Section 3 is the main part
of this text and
we shall discuss the reality conditions
for $q=\pm\overline{r}$.

\section[Preliminaries]{Preliminaries}

In this section we first recall some basic facts from the theory of symmetric Riemann surfaces
\cite{01,nata}
and
main results in our previous work \cite{05}.

\subsection{Basic facts on Riemann surface}
A Riemann surface $\cur$ of genus $n$ is called real if it admits
an antiholomorphic involution $\tau:\cur\rightarrow \cur, \,\,\tau^2=\textrm{id}|_{\cur}.$
Let $\mathcal{R}=\{P\in\cur|\tau(P)=P\}$ be the set
of fixed points of $\tau.$ The connected components of $\mathcal{R}$
are called the real ovals of $\tau.$
The number $r$ of the real ovals satisfies the relations $0\leq r\leq n+1.$
If $r=n+1,$ then $\cur$
is called an $\textrm{M}$-curve. Moreover, the set $\cur\backslash\mathcal{R}$
has either one or two connected components. We call $\cur$
is a dividing curve if $\cur\backslash\mathcal{R}$ has two components
and nondividing if  $\cur\backslash\mathcal{R}$ is connected. Obviously, an M-curve
is always a dividing curve.

  We fix a canonical homology basis $\{a_j, b_j\}_{j=1}^g$ on $\cur$
  in such a way that the intersection matrix of the
        cycles satisfies
       \begin{equation}a_j \circ b_k =\delta_{j,k},\quad a_j \circ a_k=0, \quad
        b_j \circ   b_k=0, \quad j,k=1,\ldots, n.
        \end{equation}
  Let $\omega_1,\ldots,\omega_j$ be the basis for holomorphic differentials on $\cur$,
  normalized with respect to the homology basis
  \begin{equation}\label{112}
    \int_{a_k} \omega_j = \delta_{j,k},\quad  j,\,k=1,\ldots,n
  \end{equation}
and the period matrix
\begin{equation*}
 \Gamma=(\Gamma_{j,k})= \left(\int_{b_k} \omega_j\right), \quad
          j,\,k=1, \ldots ,n.
\end{equation*}
Associated with $\Gamma$
        one defines the period lattice $L_n$ in $\mathbb{C}^n$
         by
       \begin{equation*}
        L_{n}=\{\underline{z}\in \mathbb{C}^{n}|
           ~\underline{z}=\underline{N}+\Gamma\underline{M},
           ~\underline{N},\underline{M}\in \mathbb{Z}^{n}\}.
       \end{equation*}
      The Riemann theta function associated with Riemann surface $\mathcal{K}_n$ and the homology basis $\{a_j,b_j\}_{j=1,\ldots,n}$
       is given by
       \begin{equation}\label{theta}
       \theta(\underline{z})=\sum_{\underline{n}\in\mathbb{Z}^n}\exp\Big(2\pi i(\underline{n},
       \underline{z})+\pi i(\underline{n},\Gamma\underline{n})\Big),~~\underline{z}\in\mathbb{C}^n,
       \end{equation}
        where $(\underline{A},\underline{B})=\sum_{j=1}^{n}\overline{A}_jB_j$ denotes the inner product in $\mathbb{C}^n.$
       Then the Jacobi variety $J(\mathcal{K}_n)$
       of $\mathcal{K}_n$ is defined by
       \begin{equation*}
         J(\mathcal{K}_n)=\mathbb{C}^n/L_n
       \end{equation*}
       and the Abel maps are defined by
       \begin{equation}\label{3.37}
       \begin{split}
        \underline{A}_{Q_0}:&\mathcal{K}_{n} \rightarrow
       J(\mathcal{K}_{n}), \\
       &
       P \mapsto \underline{A}_{Q_0} (P)
       =\left(\int_{Q_0}^P\omega_1,\ldots,\int_{Q_0}^P\omega_{n}\right)
      \equiv(\underline{A}_{Q_0,1}(P),\ldots,
       \underline{A}_{Q_0,n} (P))
      \end{split}
      \end{equation}
      and
       \begin{equation}\label{3.38}
       \begin{split}
        \underline{\alpha}_{Q_0}:&
       \mathrm{Div}(\mathcal{K}_{n}) \rightarrow
       J(\mathcal{K}_{n}),\\
      & ~~~~~~~~~\mathcal{D} \mapsto \underline{\alpha}_{Q_0}
       (\mathcal{D})= \sum_{P\in \mathcal{K}_{n}}
       \mathcal{D}(P)\underline{A}_{Q_0} (P) \equiv(\alpha_{Q_0,1}(\mathcal{D}),\ldots,\alpha_{Q_0,n}(\mathcal{D})),
      \end{split}
       \end{equation}
       where $Q_0$ is a fixed base point
       and the same path is chosen from $Q_0$ to $P$
       in (\ref{3.37}) and (\ref{3.38}).

\begin{prop}\label{prop1}\emph{(Vinnikov, \cite{01})}
There exists a canonical homology
basis $\{a_j,b_j\}_{j=1}^n$ ("Vinnikov basis") on $\cur$ such that
the action of $\tau$ on the homology group $\textrm{H}_1(\cur,\mathbb{Z})$
is
\begin{align}\label{2.1}
  &\left(
    \begin{array}{c}
      \tau(\underline{a}) \\
      \tau(\underline{b}) \\
    \end{array}
  \right)
  =\left(
     \begin{array}{cc}
       I_n & 0 \\
       H & -I_n \\
     \end{array}
   \right)\left(
            \begin{array}{c}
              \underline{a} \\
              \underline{b} \\
            \end{array}
          \right),
  \end{align}
  where
  \begin{align}
   &\underline{a}=(a_1,\ldots,a_n),\quad \underline{b}=(b_1,\ldots,b_n),\nonumber\\
   &\tau(\underline{a})=(\tau(a_1),\ldots,\tau(a_n)),\nonumber\\
   &\tau(\underline{b})=(\tau(b_1),\ldots,\tau(b_n)),\nonumber
\end{align}
and
$H$ is a block diagnonal $n\times n$ matrix defined as follows\\
1) If $\mathcal{R}\neq \emptyset,$
 \begin{align*}
 H=\left(
     \begin{array}{cccccc}
       \sigma_1 & &  &  &  &  \\
        & \ddots &  &  &  &  \\
       & & \sigma_1 &  & &  \\
        &  &  & 0 &  &  \\
        &  &  &  & \ddots &  \\
        & &  &  &  & 0 \\
     \end{array}
   \right)\quad \textrm{if $\mathcal{R}$ is dividing,}
\end{align*}
 \begin{align*}
  H=\left(
      \begin{array}{cccccc}
        1 &   &   &  &   &   \\
          & \ddots &   &   &   &   \\
          &   &  1 &   &   &   \\
          &   &   & 0  &   &   \\
          &   &   &  & \ddots &   \\
         &   &  &   &   & 0 \\
      \end{array}
    \right)\quad \textrm{if $\mathcal{R}$ is nondividing,}
 \end{align*}
(rank$H=n+1-r$).
\\
2) If $\mathcal{R}=\emptyset$,
\begin{align*}
   \left(
      \begin{array}{ccc}
                  \sigma_1 &   &   \\
                    & \ddots &   \\
                    &   & \sigma_1 \\
     \end{array}
   \right)\quad \textrm{or}\quad
   H=\left(
       \begin{array}{cccc}
         \sigma_1 &  &  &  \\
          & \ddots &  &  \\
          &  & \sigma_1 &  \\
          &  & & 0 \\
       \end{array}
     \right),
\end{align*}
(rank $H=n$ if $n$ is even, rank $H=g-1$ if $g$ is odd),
       \begin{equation*}
      \sigma_1=\left(
                 \begin{array}{cc}
                   0 & 1 \\
                   1 & 0 \\
                 \end{array}
               \right).
      \end{equation*}
\end{prop}

\begin{prop}\emph{(\cite{01}, Proposition 2.3)}
The Riemann theta function (\ref{theta}) associated with
the homology basis chosen in {\bf Proposition \ref{prop1}} satisfies
  \begin{flalign}
       &\overline{\theta(\bar{\underline{z}})}=c\theta(\underline{z}+\frac{1}{2}\textrm{diag}(H)),\quad c\in\mathbb{C},\quad |c|=1,\quad\underline{z}\in\mathbb{C}^n,\label{z1}\\
       &\textrm{(the constant $c$ is independent of $\Gamma$, but may dependent on H)}\nonumber\\
       &\overline{\underline{\Xi}}_{Q_0}=\underline{\Xi}_{Q_0}+\frac{1}{2}
       \textrm{diag}(H)+(n-1)\underline{\alpha}_{Q_0}(\tau(Q_0)).
  \end{flalign}
\end{prop}

For a Riemann matrix $\Gamma$ associated with a general
homology basis $\{a_j,b_j\}_{j=1}^n$, we have the following result.
\begin{prop}\label{prop113}
If $\overline{\Gamma}=\Lambda-\Gamma,$ where $\Lambda=(\Lambda_{ij}), \Lambda_{ij}\in\mathbb{Z},$
then $\overline{\theta}(\overline{\underline{z}})=\overline{\theta}(\underline{z}+\frac{1}{2}\textrm{diag}(\Lambda)).$
\end{prop}

\proof The proof follows from the definition of the function $\theta(\underline{z})$.\qed

\subsection{Algebro-geometric solutions of FL hierarchy}

 In this text, the Forkas-Lenells type nonsingular curve $\cur$ \cite{05} reads
\begin{align}\label{2.3}
   \mathcal{K}_n: &\,
    \mathcal{F}(\xi,y)=y^2-\prod_{m=0}^{2n+1}(\xi-E_m)=0,
 \end{align}
where
 $
  \,E_m\neq E_{m^\prime}\,\,
  \, \textrm{for}\,\,m\neq m^\prime,\,\,m,m^\prime=0,1,\cdots,2n+1,
    \{E_m\}_{m=0}^{2 n+1}
    \subset\mathbb{C}\backslash\{0\},\,  $
and
$ n=n_++n_--1\in\mathbb{N}, \,\, n_+,n_-\in\mathbb{N}_0.$
Based on the fundamental Riemann surface theory, $\mathcal{K}_n$ is compactified by joining two points at infinity
$P_{\infty_\pm}$, $P_{\infty_+} \neq P_{\infty_-}$, but for
notational simplicity the compactification is also denoted by
$\mathcal{K}_n$. Points $P$ on
      $$\mathcal{K}_{n} \setminus \{P_{\infty_+},P_{\infty_-}\}$$
are represented as pairs $P=(\xi,y(P))$, where $y(\cdot)$ is the
meromorphic function on $\mathcal{K}_{n}$ satisfying
       $$\mathcal{F}_n(\xi,y(P))=0.$$
For convenience we use the notation $\widehat{\xi}^{\pm}=(\xi,\pm \sqrt{\prod_{m=0}^{2n+1}(\xi-E_m)})$
by introducing an appropriate choice of the square root branch. Especially,
a branch point $P=(\xi,0)\in\cur$ is then denoted by $\widehat{\xi}.$
The complex structure on $\mathcal{K}_{n}$ is defined in the usual
way by introducing local coordinates
$$\zeta_{Q_0}:P\rightarrow(\xi-\xi_0)$$
near points $Q_0=(\xi_0,y(Q_0))\in \mathcal{K}_{n},$ which are neither branch nor singular points of
$\mathcal{K}_{n}$;
near the points $P_{\infty_\pm} \in \mathcal{K}_{n}$, the local
coordinates are
   $$\zeta_{P_{\infty_\pm}}:P \rightarrow \xi^{-1},$$
and similarly at branch and singular points of $\mathcal{K}_{n}.$
Hence $\mathcal{K}_n$ becomes a two-sheeted Riemann surface of
 topological genus $n$ in a standard manner.

The holomorphic differentials $\eta_\ell(P)$ on
         $\mathcal{K}_{n}$ are defined by
        \begin{equation}\label{3.34}
         \eta_\ell(P)= \frac{\xi^{\ell-1}}{y(P)} d\xi,
         \qquad \ell=1,\ldots,n,
        \end{equation}
 Associated with $\mathcal{K}_n$ one
        introduces an invertible
        matrix $E \in \textrm{GL}(n, \mathbb{C})$
         \begin{equation}\label{3.35}
          \begin{split}
          & E=(E_{j,k})_{n \times n}, \quad E_{j,k}=
           \int_{a_k} \eta_j, \\
          &  \underline{c}(k)=(c_1(k),\ldots, c_{n}(k)), \quad
           c_j(k)=(E^{-1})_{j,k},
           \end{split}
         \end{equation}
        and the normalized holomorphic differentials
        \begin{equation}\label{3.36}
          \omega_j= \sum_{\ell=1}^{n} c_j(\ell)\eta_\ell, \quad
          \int_{a_k} \omega_j = \delta_{j,k}, \quad
          \int_{b_k} \omega_j= \Gamma_{j,k}, \quad
          j,k=1, \ldots ,n.
        \end{equation}
        Apparently, the Riemann matrix $\Gamma=(\Gamma_{i,j})$ is symmetric and has a
        positive-definite imaginary part.
Let
 $\eta\in\mathbb{C},\,\,|\eta|<\textrm{min}\{|E_0|^{-1},|E_{1}|^{-1},|E_{2}|^{-1},\dotsi,|E_{2n+1}|^{-1}\}$
 and abbreviate
 $\underline{E}=(E_{0},\dotsi,E_{2n+1}),\,\,\underline{E}^{-1}=(E_{0}^{-1},\dotsi,E_{2n+1}^{-1}).$
      It turns out that \cite{15} \eql{ap2000}{\left(\prod_{m=0}^{2n+1}(1-E_{m}\eta)\right)^{-1/2}=
      \sum_{k=0}^{\infty}\hat{c}_{k}(\underline{E})\eta^{k},}
where $\hat{c}_{0}(\underline{E})=1,\quad \hat{c}_{1}(\underline{E})=\frac{1}{2}\sum_{m=0}^{2n+1}E_{m},$
and
$$\hat{c}_k(\underline{E})=
\sum_{\begin{smallmatrix}j_0,\dotsi,j_{2n+1}=0\\j_0+\dotsi+ j_{2n+1}=k\end{smallmatrix}}^{k}\frac{(2j_0)!\dotsi(2j_{2n+1})! }{2^{2k}(j_0!)^2(j_{2n+1}!)^2}E_0^{j_0}\ldots E_{2n+1}^{j_{2n+1}},\quad k\in\mathbb{N}.\quad$$
Similarly, one derives
 $$\left(\prod_{m=0}^{2n+1}(1-E_{m}\eta)\right)^{1/2}=\sum_{k=0}^{+\infty}{c}_{k}(\underline{E})\eta^{k},$$
where ${c}_{0}(\underline{E})=1,\quad {c}_{1}(\underline{E})=-\frac{1}{2}\sum_{m=0}^{2n+1}E_{m},$ and \eqnarrayn{
c_k(\underline{E})=
\sum_{\begin{smallmatrix}j_0,\dotsi,j_{2n+1}=0\\j_0+\dotsi+ j_{2n+1}=k\end{smallmatrix}}^{k}\frac{(2j_0)!\dotsi(2j_{2n+1})! E_0^{j_0}\ldots E_{2n+1}^{j_{2n+1}}}{2^{2k}(j_0!)^2\ldots (j_{2n+1}!)^2(2j_0-1)\dotsi (2j_{2n+1}-1)},\,\, k\in\mathbb{N}. }\vspace{0.4cm}
The Abel differentials $\Omega_{P_{0,\pm},s-1}^{(2)},\Omega_{P_{\infty\pm},s-1}^{(2)}$ of second kind are
 defined by
\begin{flalign}
\Omega_{P_{\infty\pm},s-1}^{(2)}=&\pm\frac{\sum_{j=0}^{s}
c_j(\underline{E})\xi^{n+s-j}}{\sqrt{\prod_{m=0}^{2n+1}(\xi-E_m)}}d\xi+\sum_{i=1}^{n}c_{j,\pm}^{(s-1)}
\omega_j,\nonumber\\
 =&\pm\frac{\prod_{j=1}^{n+s}(\xi-m_j^s)}{\sqrt{\prod_{m=0}^{2n+1}(\xi-E_m)}}d\xi,\label{1226a}\\
\Omega_{P_{0,\pm},s-1}^{(2)}=&\pm\frac{(-1)^{n+1}\prod_{m=0}^{2n+1}E_m^{1/2}\sum_{j=0}^{s}c_j(\underline{E})
\xi^{-(s+1)+j}}{\sqrt{\prod_{m=0}^{2n+1}(\xi-E_m)}}d\xi\nonumber\\
&+\sum_{i=1}^{n}\tilde{c}_{j,\pm}^{(s-1)}\omega_j,\nonumber\\
 =&\pm\frac{\xi^{-(s+1)}\prod_{j=1}^{n+s}(\xi-\widetilde{m}_j^s)}{\sqrt{\prod_{m=0}^{2n+1}(\xi-E_m)}}d\xi,
 \label{1227a}
\end{flalign}
where the constants $c_{j,\pm}^{(s-1)},\tilde{c}_{j,\pm}^{(s-1)},m_j^s,\widetilde{m}_j^s$ are determined by the normalization
conditions
\begin{flalign}
 \int_{a_j}\Omega_{P_{\infty\pm},s-1}^{(2)}=0,\quad j=1,\ldots,n,\label{20136901}\\
 \int_{a_j}\Omega_{P_{0,\pm},s-1}^{(2)}=0,\quad j=1,\ldots,n.\label{20136902}
\end{flalign}
By introducing
\begin{align}
  \widetilde{\Omega}_{\underline{r}}^{(2)}=& \Big(\sum_{s=1}^{r_-}s\tilde{c}_{r_--s,-}
  (\Omega_{{P_{0,+},s-1}}^{(2)}-\Omega_{{P_{0,-}},s-1}^{(2)})\nonumber\\
  &+\sum_{s=1}^{r_+}s\tilde{c}_{r_+-s,+}
  (\Omega_{{P_{\infty+}},s-1}^{(2)}-\Omega_{{P_{\infty-}},s-1}^{(2)})\Big)\label{a2.28}\\
  \widetilde{\Omega}_{\underline{r}}^{0,\pm}=&\lim_{P\rightarrow P_{0,\pm}}\Big(\int_{Q_0}^P\widetilde{\Omega}_{\underline{r}}^{(2)}\mp
  \sum_{s=1}^{r_-}\tilde{c}_{r_--s}\zeta^{-s}\Big),\,\,\tilde{c}_{s}\in\mathbb{R},\\
  \Omega_0^{(2)}=&\Omega_{P_{\infty+},0}^{(2)}-\Omega_{P_{\infty-},0}^{(2)},\\
  \int_{Q_0}^P\Omega_0^{(2)}&\underset{\zeta\rightarrow 0}{=}
                  e_{0,\pm}+e_{1,\pm}\zeta+O(\zeta^2),\quad P\rightarrow P_{0,\pm},\\
   \underline{U}_0^{(2)}=&(\underline{U}_{0,1}^{(2)},\ldots,\underline{U}_{0,n}^{(2)}),
                 \quad \underline{U}_{0,j}^{(2)}=\frac{1}{2\pi i}\int_{b_j}\Omega_0^{(2)},\quad j=1,\ldots,n,\\
    \widetilde{\underline{U}}_{\underline{r}}^{(2)}
  =&( \widetilde{\underline{U}}_{\underline{r},1}^{(2)},\ldots, \widetilde{\underline{U}}_{\underline{r},n}^{(2)}),\quad  \widetilde{\underline{U}}_{\underline{r},j}^{(2)}=\frac{1}{2\pi i}\int_{b_j}\widetilde{\Omega}_{\underline{r}}^{(2)},\quad j=1,\ldots,n,\label{a2.33}
\end{align}
the solution of the initial problem \cite{05}
\begin{flalign}\label{a6.6}
  \textrm{$\widetilde{\text{FL}}$}_{\underline{r}}(q,r)=0,\quad &(q,r)|_{t_{\underline{r}}=t_{0,\underline{r}}}=(q^{(0)},r^{(0)})
  \end{flalign}
 has the form
 \begin{align}
        q(x,t_{\underline{r}})=& q(x_0,t_{0,\underline{r}})\frac{\theta(\underline{\xi}(P_{0,-},
        \hat{\underline{\mu}}(x_0,t_{0,\underline{r}})))}
       {\theta(\underline{\xi}(P_{0,-},\hat{\underline{\mu}}(x,t_{\underline{r}})))}
       \frac{\theta(\underline{\xi}(P_{0,+},\hat{\underline{\mu}}(x,t_{\underline{r}})))}
       {\theta(\underline{\xi}(P_{0,+},\hat{\underline{\mu}}(x_0,t_{0,\underline{r}})))} \nonumber\\
       &~~~~~~~~~~~~~~~~ \times
       \exp \Big(i(x-x_0)(e_{0,-}-e_{0,+})+i(t_{\underline{r}}-t_{0,\underline{r}})
       (\widetilde{\Omega}_{\underline{r}}^{0,-}-\widetilde{\Omega}_{\underline{r}}^{0,+})\Big),
       \label{solutionqt}\\
        r(x,t_{\underline{r}})= &r(x_0,t_{0,\underline{r}})
       \frac{\theta(\underline{\xi}(P_{0,-},\hat{\underline{\nu}}(x,t_{\underline{r}})))}
       {\theta(\underline{\xi}(P_{0,-},\hat{\underline{\nu}}(x_0,t_{0,\underline{r}})))}
       \frac{\theta(\underline{\xi}(P_{0,+},\hat{\underline{\nu}}(x_0,t_{0,\underline{r}})))}
       {\theta(\underline{\xi}(P_{0,+},\hat{\underline{\nu}}(x,t_{\underline{r}})))}\nonumber\\
       &~~~~~~~~~~~\times \exp\Big(-i(x-x_0)(e_{0,-}-e_{0,+})-i(t_{\underline{r}}-t_{0,\underline{r}})
       (\widetilde{\Omega}_{\underline{r}}^{0,-}+\widetilde{\Omega}_{\underline{r}}^{0,+})\Big),
       \label{solutionrt}\\
         q(x_0,t_{0,\underline{r}})&r(x_0,t_{0,\underline{r}})=
       \frac{\theta(\underline{\xi}(P_{0,-},\hat{\underline{\nu}}(x_0,t_{0,\underline{r}})))}
       {\theta(\underline{\xi}(P_{0,-},\hat{\underline{\mu}}(x_0,t_{0,\underline{r}})))}
       \frac{\theta(\underline{\xi}(P_{0,+},\hat{\underline{\mu}}(x_0,t_{0,\underline{r}})))}
       {\theta(\underline{\xi}(P_{0,+},\hat{\underline{\nu}}(x_0,t_{0,\underline{r}})))}\nonumber\\
        &\times \exp\Big(\omega_0^{0,-}-\omega_{0}^{0,+}\Big),\label{qrt}
      \end{align}
  where we use the abbreviations
            \begin{eqnarray}\label{3.46}
               &&
           \underline{\xi}(P,\underline{Q})= \underline{\Xi}_{Q_0}
           -\underline{A}_{Q_0}(P)+\underline{\alpha}_{Q_0}
             (\mathcal{D}_{\underline{Q}}), \nonumber \\
          &&
           P\in \mathcal{K}_{n},\,
          \underline{Q}=(Q_1,\ldots,Q_{n})\in
          \mathrm{Sym}^{n}(\mathcal{K}_{n}),
         \end{eqnarray}
and $\underline{\Xi}_{Q_0}$ is
the vector of Riemann constants (cf.(A.45) \cite{15}). Moreover,
The Abel map linearizes the auxiliary divisors
       $\mathcal{D}_{\hat{\underline{\mu}}(x,t_{\underline{r}})},
       \mathcal{D}_{\hat{\underline{\nu}}(x,t_{\underline{r}})}$
       in the sense that
         \begin{align}
         \underline{\alpha}_{Q_0}(\mathcal{D}_{\underline{\hat{\mu}}(x,t_{\underline{r}})})
         =&\underline{\alpha}_{Q_0}(\mathcal{D}_{\underline{\hat{\mu}}(x_0,t_{0,\underline{r}})})
         -i\underline{U}_0^{(2)}(x-x_0)-i\widetilde{\underline{U}}_{\underline{r}}^{(2)}(t-t_{0,\underline{r}}),\label{3.52AA}\\
         \underline{\alpha}_{Q_0}(\mathcal{D}_{\underline{\hat{\nu}}(x,t_{\underline{r}})})
         =&\underline{\alpha}_{Q_0}(\mathcal{D}_{\underline{\hat{\nu}}(x_0,t_{0,\underline{r}})})
         -i\underline{U}_0^{(2)}(x-x_0)-i\widetilde{\underline{U}}_{\underline{r}}^{(2)}(t-t_{0,\underline{r}}).\label{3.52BB}
       \end{align}
       For convenience we
       denote by
       \begin{flalign*}
         \textbf{Z}=&\underline{\Xi}_{Q_0}-\underline{A}_{Q_0}(P_{0,-})+
         \underline{\alpha}_{Q_0}(\mathcal{D}_{\underline{\hat{\mu}}(0,0)}),\\
         \textbf{Y}=&\underline{\Xi}_{Q_0}-\underline{A}_{Q_0}(P_{0,-})+
         \underline{\alpha}_{Q_0}(\mathcal{D}_{\underline{\hat{\nu}}(0,0)})\\
        \textbf{T}=&\underline{A}_{P_{0,-}}(P_{0,+}),\quad
         \textbf{V}=-i\underline{U}_0^{(2)},\quad \textbf{W}=-i\widetilde{\underline{U}}_{\underline{r}}^{(2)}
       \end{flalign*}
       and then
       (\ref{solutionqt}), (\ref{solutionrt})
       can be rewritten as (taking $(x_0,t_{0,\underline{r}})=0$)
       \begin{flalign}
         q(x,t_{\underline{r}})=&q(0,0)\frac{\theta(\textbf{Z})}
          {\theta(\textbf{Z}-\textbf{T})}\frac{\theta(\textbf{Z}-\textbf{T}
          +\textbf{V}x+\textbf{W}t_{\underline{r}})}{\theta(\textbf{Z}
          +\textbf{V}x+\textbf{W}t_{\underline{r}})}\nonumber\\
         &\times\exp\left(i(e_{0,-}-e_{0,+})x+i
       (\widetilde{\Omega}_{\underline{r}}^{0,-}-
       \widetilde{\Omega}_{\underline{r}}^{0,+})t_{\underline{r}}\right),\label{1223}\\
       r(x,t_{\underline{r}})=&r(0,0)\frac{\theta(\textbf{Y}-\textbf{T})}
          {\theta(\textbf{Y})}\frac{\theta(\textbf{Y}
          +\textbf{V}x+\textbf{W}t_{\underline{r}})}{\theta(\textbf{Y}-\textbf{T}
          +\textbf{V}x+\textbf{W}t_{\underline{r}})}\nonumber\\
         &\times\exp\left(-i(e_{0,-}-e_{0,+})x-i
       (\widetilde{\Omega}_{\underline{r}}^{0,-}
       -\widetilde{\Omega}_{\underline{r}}^{0,+})t_{\underline{r}}\right),\label{1224}\\
       q(0,0)r(0,0)&=\frac{\theta(\textbf{Y})}
          {\theta(\textbf{Z})}\frac{\theta(\textbf{Z}-\textbf{T})}
          {\theta(\textbf{Y}-\textbf{T})}\label{1225},
       \end{flalign}
       that is,
       \begin{flalign}
         q(x,t_{\underline{r}})=&q(0,0)\frac{\theta(\textbf{Z})}
          {\theta(\textbf{Z}-\textbf{T})}\frac{\theta(\textbf{Z}-\textbf{T}
          +\textbf{V}x+\textbf{W}t_{\underline{r}})}{\theta(\textbf{Z}
          +\textbf{V}x+\textbf{W}t_{\underline{r}})}\nonumber\\
         &\times\exp\left(i(e_{0,-}-e_{0,+})x+i
       (\widetilde{\Omega}_{\underline{r}}^{0,-}-
       \widetilde{\Omega}_{\underline{r}}^{0,+})t_{\underline{r}}\right),\label{1227}\\
       r(x,t_{\underline{r}})=&\frac{1}{q(0,0)}\frac{\theta(\textbf{Y}-\textbf{T})}
          {\theta(\textbf{Z})}\frac{\theta(\textbf{Y}
          +\textbf{V}x+\textbf{W}t_{\underline{r}})}{\theta(\textbf{Y}-\textbf{T}
          +\textbf{V}x+\textbf{W}t_{\underline{r}})}\nonumber\\
         &\times\exp\left(-i(e_{0,-}-e_{0,+})x-i
       (\widetilde{\Omega}_{\underline{r}}^{0,-}
       -\widetilde{\Omega}_{\underline{r}}^{0,+})t_{\underline{r}}\right).\label{1228}
\end{flalign}
It should be noticed that
the following relation
\begin{flalign}\label{1226}
  \textbf{Z}=\textbf{Y}-\textbf{T}-\textbf{Q},\quad
  \textbf{Q}=\underline{A}_{P_{0,+}}(P_{\infty+})
\end{flalign}
holds
by the equivalence of the divisors
$\mathcal{D}_{P_{0,-}\underline{\hat{\nu}}(x,t_{\underline{r}})}\sim
\mathcal{D}_{P_{\infty+}\underline{\hat{\mu}}(x,t_{\underline{r}})}$.

\section{Two results about the spectral data}

In this section, we give two
results about the spectral data
\begin{equation}\label{spectraldata}\{\hat{\mu}_j(0,0),\hat{\nu}_j
(0,0),E_m;q(0,0),
r(0,0)\}_{\begin{smallmatrix}j=1,\ldots,n,m=0,\ldots,2n+1\end{smallmatrix}}.\end{equation}
The first theorem (cf. \textbf {Theorem \ref{thm201}})
shows any smooth solution in $\mathcal{M}^{\underline{n},\underline{r}}_0$
can be locally and uniquely determined by the spectral data.
And the inverse is also true.
Therefore, the reality conditions for
$q=\pm\bar{r}$
are totally reflected
on the spectral parameters. The
\textbf {Theorem \ref{thm203}} can be regarded as
a complementary illustration that all the theta functions
appeared
in this paper is not identical to zero.

In this text, a function $p=p(x,t_{\underline{r}})$ in two real variables
$x,t_{\underline{r}}$
is called proper at $(0,0)$
if there exists an open interval $\Omega\in\mathbb{R}^2$ with $(0,0)\in\Omega$
such that $p$ is smooth at all points $x\in\Omega$.
Keeping $n,\underline{r}$ fixed, we denote by $\mathcal{M}^{\underline{n},\underline{r}}$
all the smooth proper solutions
of the Forkas-Lenells initial value problem (\ref{a6.6}) at $(0,0)$.
The subset $\mathcal{M}^{\underline{n},\underline{r}}_0
\subset\mathcal{M}^{\underline{n},\underline{r}}$
is a collection of smooth solutions in $\mathcal{M}^{\underline{n},\underline{r}}$
such that
the polynomials $F_{\underline{n}}(x,t_{\underline{r}};\xi), H_{\underline{n}}(x,t_{\underline{r}};\xi)$
in $\xi$
only have simple roots at $(x, t_{\underline{r}})=(0,0),$ respectively.

\begin{thm}\label{thm201}
Assume $q, r\in C^\infty(\Omega)$
on some open set $\Omega\subset\mathbb{R}^2$ with $(0,0)\subset\Omega$. Moreover,
assume $\mu_j(0,0)\neq \mu_k(0,0), \nu_j(0,0)\neq \nu_k(0,0),
$ and $\mu_j(0,0)\neq 0, \nu_j(0,0)\neq 0$
for $j\neq k, j,k=1,\ldots,n.$
Then there exists a one-to-one correspondence
between
the spectral data $(\ref{spectraldata})$ and $\mathcal{M}^{\underline{n},\underline{r}}_0$
\begin{flalign}\label{a2.021}
&\{\hat{\mu}_j(0,0),\hat{\nu}_j
(0,0),E_m;q(0,0),
r(0,0)\}_{\begin{smallmatrix}j=1,\ldots,n,m=0,\ldots,2n+1\end{smallmatrix}}
\rightleftharpoons  \mathcal{M}^{\underline{n},\underline{r}}_0(\Omega).
\end{flalign}

\end{thm}
\proof
We first prove (\ref{a2.021})
holds for stationary case, that is,
there exists an open interval $\mathcal{I}_1\subset\mathbb{C},$
$0\in\mathcal{I}_1$ such that
the correspondence
\begin{flalign*}
 \{\mu_j(0),\nu_j(0),E_m;q(0),r(0)
 \}_{ j=1,\ldots,n, m=0,\ldots,2n+1 }\rightleftharpoons\{q(x),r(x)\}
\end{flalign*}
is one to one on $\mathcal{I}_1.$ It suffices
to build the
relation from the algebro-geometric data $\{\mu_j(x_0),\nu_j(x_0),E_m;q(x_0),r(x_0)
 \}_{ j=1,\ldots,n, m=0,\ldots,2n+1 }$ to stationary solutions $q(x),r(x)$
 since another direction is obvious.
 The Dubrovin-type equations (\cite{05}, Lemma 3.2)
\begin{flalign}
 \mu_{j,x}(x)  =&\frac{-2iy(\hat{\mu}_j(x))}{\prod_{k=1,k\neq j}^{n}(\mu_j(x)-\mu_k(x))},\ \ j=1,\ldots,n,\\
 \nu_{j,x}(x)  =&\frac{-2iy(\hat{\mu}_j(x))}{\prod_{k=1,k\neq j}^{n}(\nu_j(x)-\nu_k(x))},\ \ j=1,\ldots,n,\,\,
\end{flalign}
has a unique solution
\begin{flalign*}
 \hat{\mu}_j \in C^\infty(\Omega_\mu^\prime,\mathcal{K}_{n}),\quad
 \hat{\nu}_j \in C^\infty(\Omega_\nu^\prime,\mathcal{K}_{n}),\quad
      \quad j=0,\ldots,n
\end{flalign*}
on some neighborhood $\Omega_\mu^\prime,\Omega_\nu^\prime\subset \mathbb{R}$ with $0\in\Omega_\mu^\prime,\Omega_\nu^\prime.$
Using
the trace formulas (\cite{05}, Theorem 3.3)
\begin{flalign}
\frac{iq}{2q_x}=(-1)^n\prod_{j=1}^n\mu_j,\quad \frac{ir}{2r_x}=(-1)^{n-1}\prod_{j=1}^n\nu_j,
\end{flalign}
one gets $q,r$ has a unique solution
\begin{flalign}
 q(x)=&q(0)\exp\Big(\int_{0}^x\frac{i}{2(-1)^n\prod_{j=1}^n\mu_j(x^\prime)}dx^\prime\Big)\in
 C^\infty(\Omega_\mu,\mathcal{K}_{n}),\label{a.2025}\\
 r(x)=&r(0)\exp\Big(\int_{0}^x\frac{i}{2(-1)^{n-1}\prod_{j=1}^n\nu_j(x^\prime)}dx^\prime\Big)\in
 C^\infty(\Omega_\nu,\mathcal{K}_{n}).\label{a.2026}
\end{flalign}
Taking $\mathcal{I}_1=\Omega_\mu^\prime \bigcap \Omega_\nu^\prime$, then we prove this Theorem for stationary case.
Analogous to time-dependent case, (\ref{a.2025}) and (\ref{a.2026})
change to
\begin{flalign}
 q(x,t_{\underline{r}})=&q(0,t_{\underline{r}})
 \exp\Big(\int_{0}^x\frac{i}{2(-1)^n\prod_{j=1}^n\mu_j(x^\prime,t_{\underline{r}})}dx^\prime\Big),\label{a2027}\\
 r(x,\underline{r})=&r(0,t_{\underline{r}})\exp\Big(\int_{0}^x\frac{i}{2(-1)^{n-1}\prod_{j=1}^n
 \nu_j(x^\prime,t_{\underline{r}})}dx^\prime\Big),\label{a2028}
\end{flalign}
and
\begin{flalign} \label{a2029}
\mu_j(x,t_{\underline{r}})\in C^{\infty}(\Omega_\mu),\quad
\nu_j(x,t_{\underline{r}})\in C^{\infty}(\Omega_\nu), \quad (0,0)\in \Omega_\mu, \Omega_\nu\subseteq\mathbb{R}^2.
\end{flalign}
In other way,
one can
prove
$f_{\ell,\pm}(0,t_{\underline{r}}), g_{\ell,\pm}(0,t_{\underline{r}}),
h_{\ell,\pm}(0,t_{\underline{r}})$
satisfying the following first order
autonomous system
\begin{flalign}
 f_{\ell,\pm,t_{\underline{r}}}(0,t_{\underline{r}})=&\mathcal{F}_{\ell, \pm}
 (f_{j,\pm}(0,t_{\underline{r}}), g_{j,\pm}(0,t_{\underline{r}}), h_{j,\pm}(0,t_{\underline{r}})),\quad \ell=0,\ldots, n_{\pm},\nonumber\\
 g_{\ell,\pm,t_{\underline{r}}}(0,t_{\underline{r}})=&\mathcal{G}_{\ell, \pm}
 (f_{j,\pm}(0,t_{\underline{r}}), g_{j,\pm}(0,t_{\underline{r}}), h_{j,\pm}(0,t_{\underline{r}})),\quad \ell=0,\ldots, n_{\pm}-\delta_{\pm},\nonumber\\
 h_{\ell,\pm,t_{\underline{r}}}(0,t_{\underline{r}})=&\mathcal{H}_{\ell, \pm}
 (f_{j,\pm}(0,t_{\underline{r}}), g_{j,\pm}(0,t_{\underline{r}}), h_{j,\pm}(0,t_{\underline{r}})),\quad \ell=0,\ldots, n_{\pm},\nonumber
\end{flalign}
where $\mathcal{F}_{\ell, \pm},\mathcal{G}_{\ell, \pm},
\mathcal{H}_{\ell, \pm}$ are polynomials
and $\delta_+=1, \delta_-=0$. This yields
the local smooth solution
$f_{\ell,\pm}(0,t_{\underline{r}}),g_{\ell,\pm}(0,t_{\underline{r}}),f_{\ell,\pm}(0,t_{\underline{r}})$
on some open interval $\mathcal{I}_2$ with $0\in\mathcal{I}_2.$ Noticing that
the Forkas-Lenells hierarchy (\cite{05}, (2.37))
is equivalent to the following
first-order system (taking $x=0$ here)
\begin{flalign}
q_{t_{\underline{r}}}(x,t_{\underline{r}})=&f_{n_-,-}(x,t_r)-f_{n_+-1,+}(x,t_{\underline{r}})+C_1,
\\
r_{t_{\underline{r}}}(x,t_{\underline{r}})=&h_{n_+-1,+}(x,t_r)-f_{n_-,-}(x,t_{\underline{r}})+C_2, \quad C_1, C_2\in\mathbb{C},
\end{flalign}
one derives
\begin{equation} \label{a2032}
q(0,t_{\underline{r}}), r(0,t_{\underline{r}})\in C^{\infty}(\mathcal{I}_2, \mathbb{R}).
\end{equation}
Finally, taking $\mathcal{I}=(\mathcal{I}_1\times\mathcal{I}_2)\bigcap \Omega_\mu \bigcap \Omega_\nu \bigcap \Omega$ and using (\ref{a2027})-(\ref{a2029}),(\ref{a2032}), we complete the proof.
\qed \\

\begin{rem}
The choice of the point (0,0) is not essential. In fact, every point
in $(x_0,t_{0,r})\in\Omega$, there exists such a kind of
structure.
\end{rem}
\begin{rem}
Due to the equivalence of the divisors
$\mathcal{D}_{P_{0,-}\underline{\hat{\nu}}(x,t_{\underline{r}})}\sim
\mathcal{D}_{P_{\infty+}\underline{\hat{\mu}}(x,t_{\underline{r}})}$ and (\ref{1225}),
the spectral data $(\ref{spectraldata})$ can be more
precisely to be written as
\begin{equation}\label{spectraldata2}
\{\hat{\mu}_j(0,0),E_m;q(0,0)\}_{\begin{smallmatrix}j=1,\ldots,n,m=0,\ldots,2n+1\end{smallmatrix}}
\end{equation}
in the algebro-geometric background \cite{15}.
\end{rem}

Following the methodology of \cite{16} handling $(1+1)$-dimensional discrete model, we shall prove the following
results.
\begin{thm}\label{thm203}
Assume $q, r\in C^\infty(\Omega)$
on some open set $\Omega\subset\mathbb{R}^2$ with $(0,0)\subset\Omega$.
Let $\mathcal{D}_{\underline{\mu}},\mathcal{D}_{\underline{\hat{\nu}}},
\underline{\hat{\mu}}=\{\hat{\mu}_1,\ldots,\hat{\mu}_n\},
\underline{\hat{\nu}}=\{\hat{\nu}_1,\ldots,\hat{\nu}_n\},$
be the pole and zero divisor of degree $n,$ respectively,
associated with $q,r$ and $\phi(\cdot,x,t_{\underline{r}})\in\cur$
according to (\cite{05}, (4.16) and (4.17)),
 \begin{flalign}
          \hat{\mu}_j(x,t_{\underline{r}})&=(\mu_j(x,t_{\underline{r}}),
          -\mu_j(x,t_{\underline{r}})^{2n_-}G_{\underline{n}}(\mu_j(x,t_{\underline{r}}),x,t_{\underline{r}})),\quad j=1,\ldots,n,\\
     \hat{\nu}_j(x,t_{\underline{r}})&=(\nu_j(x,t_{\underline{r}}),\nu_j(x,t_{\underline{r}})^{2n_-}
          G_{\underline{n}}(\nu_j(x,t_{\underline{r}}),x,t_{\underline{r}})),\quad j=1,\ldots,n.
   \end{flalign}
Then $\hat{\mu}_j(x,t_{\underline{r}})$
and $\hat{\nu}_j(x,t_{\underline{r}})$
are nonspecial for all $(x,t_{\underline{r}})\in\Omega$.
\end{thm}
\proof
The divisor
$\mathcal{D}_{\underline{\hat{\mu}}(x,t_{\underline{r}})}$ is nonspecial if and
only if $\{\hat{\mu}_1(x,t_{\underline{r}}),\dotsi,\hat{\mu}_n(x,t_{\underline{r}})\}$ contains one
pair of $\{\hat{\mu}_j(x,t_{\underline{r}}),\hat{\mu}^*_j(x,t_{\underline{r}})\}$. Hence,
$\mathcal{D}_{\underline{\hat{\mu}}(x,t_{\underline{r}})}$
is nonspecial as long as the projection $\mu_j$ of $\hat{\mu}_j$ are mutually distinct,
$\mu_j(x,t_{\underline{r}})\neq\mu_k(x,t_{\underline{r}})$ for $j\neq k$. If two or more projection coincide for some $(x_0,t_{0,\underline{r}})\in\mathbb{R}^2$,
for instance,$$\mu_{j_1}(x_0,t_{0,\underline{r}})=\ldots=
\mu_{j_k}(x_0,t_{0,\underline{r}})=\mu_0,\quad k>1,$$
then there are two cases in the
following associated with $\mu_0$.\\
(i) $\mu_0\notin \{E_0, E_1,\dotsi,E_{2p+1}\}$;
we have $G_{\underline{n}}(\mu_0,x_0,t_{0,\underline{r}})\neq 0$ and
$\hat{\mu}_{j_1}(n_0),\ldots,$ $\hat{\mu}_{j_k}(n_0)$ all meet in the same
sheet. Hence no special divisor
can arise in this manner. \\
(ii)$\mu_0\in \{E_0, E_1,\dotsi,E_{2p+1}\}$;
We assume $\mu_0=E_0$ without loss of generality.
One concludes
$F_{\underline{n}}(\mu_0,x_0,t_{0,\underline{r}})\underset{\xi\rightarrow E_{0}}{=}O\left((\xi-E_{0})^2\right)$
and $G_{\underline{n}}(\mu_0,x_0,t_{0,\underline{r}})=0$.
Hence
 \[R_{\underline{n}}(\xi,x_0,t_{0,\underline{r}})=G_{\underline{n}}^2(\xi,x_0,t_{0,\underline{r}}) +F_{\underline{n}}(\xi,x_0,t_{0,\underline{r}})H_{\underline{n}}(\xi,x_0,t_{0,\underline{r}})
 =O\left((\xi-E_{0})^2\right).\]
This conclusion contradict with the hypothesis (\ref{2.3})
that the curve is nonsingular.
As a result, we have $k=1$ and  $\hat{\mu}_j,~j=1,\ldots,n$ are pairwise distinct.
Then we have completed the proof.

\qed

\section{Reality conditions for $q=\pm\overline{r}$}

The solution (\ref{1223}), (\ref{1224}),
which does not involve a specific basis of cycles, did not lead to a expression for
the reality condition. As we know now, a formula for the reality
condition can only be written in a very specific basis.
In this section, we impose some constraints on $\cur$
and then study the parameters in
the algebro-geometric solutions of FL hierarchy.


First we consider the simplest case that all the
the branch points $\{E_m\}_{m=0}^{2n+1}$ of $\cur$ are real, that is,
 \begin{equation}\label{311}
   E_m\in\mathbb{R},\quad m=0,\ldots,2n+1
 \end{equation}
 and
 $\{E_m\}_{m=0}^{2n+1}$ are ordered such that
  \begin{equation}\label{312}
    E_0<\ldots<E_{2n+1}, \quad.
  \end{equation}

For convenience we denote by $\cur^1$ the curve $\cur$ (\ref{2.3}) under the constraint (\ref{311}) and (\ref{312}). One defines
the antiholomorphic involution on $\cur^1$
\begin{equation}\label{313}
\tau_1 (\xi,y)=(\bar{\xi},\bar{y}),
\end{equation}
where $\bar{\xi},\bar{y}$ is the complex conjugate of $\xi,y\in\mathbb{C},$ respectively.
Here we choose the square
root in
 $\sqrt{\prod_{m=0}^{2n+1}(\xi-E_m)}$ such that
 $\overline{\sqrt{\prod_{m=0}^{2n+1}(\xi-E_m)}}=\sqrt{\prod_{m=0}^{2n+1}(\xi-E_m)}$
 for $ \xi\in\mathbb{C}\backslash\bigcup_{j=0}^n[E_{2j,E_{2j+1}}].$
 Moreover, the homology basis is explicitly shown in Figure 1.
\begin{figure}[htbp]
       \begin{center}
       \includegraphics{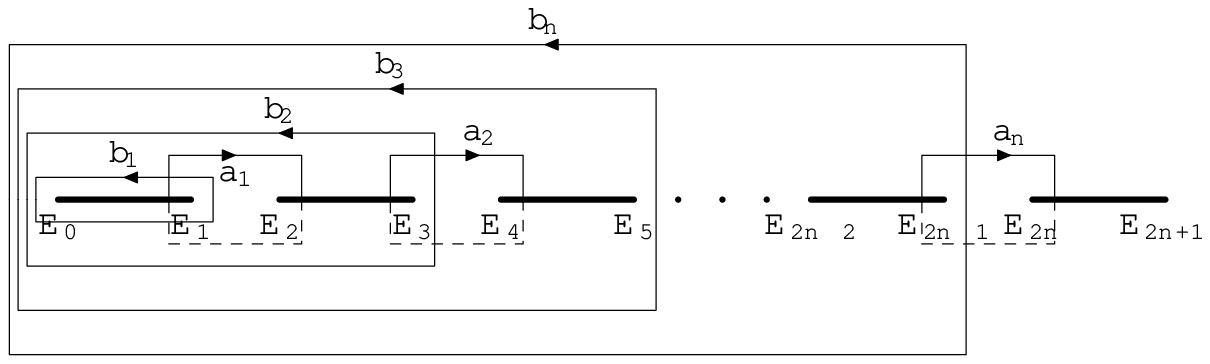}
      \caption{Homology basis on the real curves $\cur^1$, contours on sheet 1 are solid, contours on sheet 2
       are dashed}
      \end{center}
      \end{figure}

\begin{figure}[htbp]
       \begin{center}
       \includegraphics{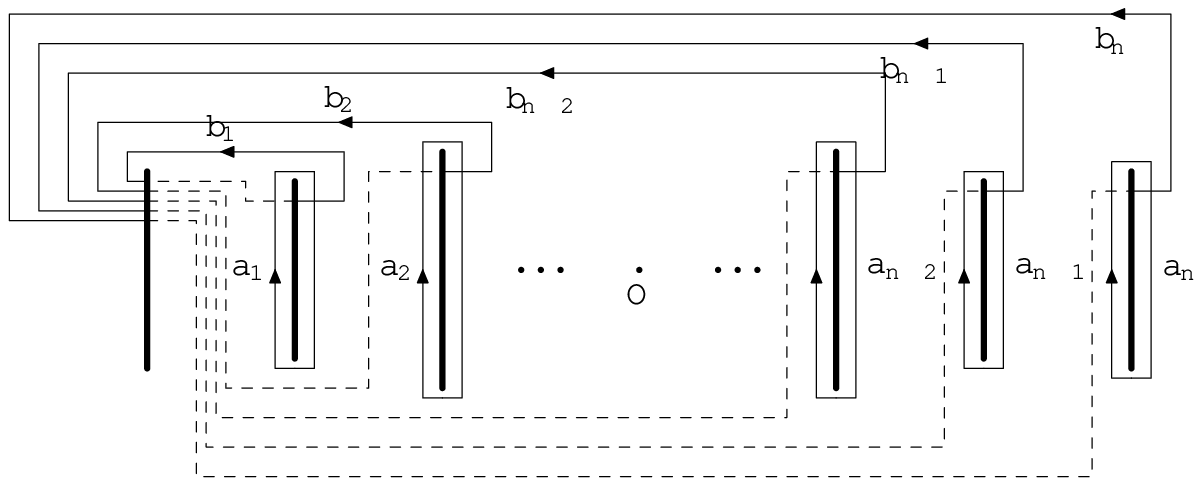}
      \caption{The path from $P_{0,+}$ to $P_{0,-}$, $\widetilde{Q}_0$ is the nearest branch point
      whose projection to $\xi$-sphere has the minimal distance to the point 0 }
      \end{center}
      \end{figure}

Obviously, the involution $\tau_1$
transforms the homology $\{a_j, b_j\}_{j=1}^n$ as follows:
\begin{equation}\label{tau1}
  \tau_1(a_j)=a_j,\,\,\tau_1(b_j)=-b_j.
\end{equation}
In other way, the real ovals of $\cur^1$ w.r.t $\tau_1$ has
$n+1$ components whose projection to $\mathbb{CP}^1$
are
\begin{equation*}
  [E_1, E_2],\,\, [E_3, E_4],\ldots, [E_{2n+1}, E_0]
\end{equation*}
and hence $\cur^1$
is an $M$-curve. This turns out that
$H=0$ in Proposition 1.1.1. Therefore,
the homology depicted in Figure 1 satisfies the condition
of Proposition 1.1.1. Furthermore, the properties of Riemann theta function
in Proposition 1.1.2 holds w.r.t the homology in Figure 1.

Next we consider the action $\tau_1$ on the Abel differentials.
Let $\tau_1^*\omega_j$ be the action of $\tau_1$ lifted to the
holomorphic differentials,
 \begin{equation}\label{2.2c}
       \tau^*_1\omega_j(P)=\omega_j(\tau_1 P),
 \end{equation}
and the definitions for the Abel differentials of the second kind and third kind
are similar.
 The normalization condition (\ref{3.36})
       indicates
\begin{flalign}\label{3050}
\int_{a_k}\overline{\tau_1^*\omega_j}=\int_{\tau_1(a_k)}\overline{\omega}_j
=\int_{a_k}\overline{\omega}_j=\delta_{j,k}=\int_{a_k}\omega_j,
\end{flalign}
and therefore
\begin{flalign}
  \int_{a_k}\Big(\overline{\tau_1^*\omega_j}-\omega_j\Big)=0,\quad j,k=1,\ldots,n.
\end{flalign}
Therefore, by the uniqueness theorem for holomorphic Abel differentials
of first kind, $\overline{\tau_1^*\omega_j}=\omega_j, j=1,\ldots,n$ and $\overline{\Gamma}=-\Gamma.$
To be more exactly, the $k$-th column vector of the matrix $\Gamma$
\begin{flalign}
 \Gamma^{(k)}=&
 \left(
   \begin{array}{c}
    \mathlarger{\int}_{b_k}\omega_1 \\
     \vdots \\
     \mathlarger{\int}_{b_k}\omega_n
   \end{array}
 \right)
 =
 \left(
   \begin{array}{c}
    \mathlarger{\int}_{\beta_k}\omega_1- \mathlarger{\int}_{\tau(\beta_k)}\omega_1 \\
     \vdots \\
     \mathlarger{\int}_{\beta_k}\omega_n- \mathlarger{\int}_{\tau(\beta_k)}\omega_n
   \end{array}
 \right)\nonumber\\
 =&2\left(
   \begin{array}{c}
    \mathlarger{\int}_{\beta_k}\omega_1 \\
     \vdots \\
     \mathlarger{\int}_{\beta_k}\omega_n
   \end{array}
 \right)
 =
 2 \left(
   \begin{array}{c}
    \sum\limits_{j=1}^{k}\mathlarger{\int}_{E_{2j-1}}^{E_{2j-2}}\omega_1 \\
     \vdots \\
     \sum\limits_{j=1}^{k}\mathlarger{\int}_{E_{2j-1}}^{E_{2j-2}}\omega_n
   \end{array}
 \right)\in i\mathbb{R}^n,\label{a2.26}
 \end{flalign}
 and
 \begin{flalign}\label{a3053}
 \int_{b_{n+1}}\omega_j=&2\int_{\beta_{n+1}}\omega_j= -\sum\limits_{k=1}^n\int_{a_k}\omega_j=-2\sum\limits_{k=1}^n\int_{E_{2k-1}}^{E_{2k}}\omega_j\in \mathbb{R},
\end{flalign}
where $\beta_k$ is an oriented curve from $\widehat{E}_{2k-1}(k=1,\ldots,n+1)$ to $\widehat{E}_{0}$.
One easily proves $\overline{\tau_1^*\eta_j}=\eta_j, j=1,\ldots,n,$
from which it follows $c_j(\ell)\in\mathbb{R},\,j,\ell=1,\ldots,n.$
Due to the normalization conditions (\ref{20136901}) and (\ref{20136902}), one easily finds
$c_{j,\pm}^{(2s-1)},\tilde{c}_{j,\pm}^{(2s-1)}\in\mathbb{R}$. Therefore,
the action of $\tau$ on $\Omega_{P_{\infty\pm},2s-1}^{(2)},\Omega_{P_{0,\pm},2s-1}^{(2)}$
is
\begin{flalign}
 \tau^*_1\Omega_{P_{\infty\pm},2s-1}^{(2)}=\overline{\Omega}_{P_{\infty\pm},2s-1}^{(2)},\label{a2.24}\\
 \tau^*_1\Omega_{P_{0,\pm},2s-1}^{(2)}=\overline{\Omega}_{P_{0,\pm},2s-1}^{(2)}.\label{a2.25}
\end{flalign}
Choose $Q_0=\widetilde{Q}_0$ (cf. Figure 2) as the base point of the Abel map
and then the following lemma
describes the properties of the parameters with respect to the constraint ($\cur^1$,$\tau_1$).
\begin{thm}\label{lem3.3.1}
Assume the algebro-geometric solution
$\{q,r\}\in\mathcal{M}^{\underline{n},\underline{r}}_0$ (see (\ref{1227}), (\ref{1228}))
and let
$\{\hat{\mu}_j(0,0), E_m;
q(0,0)\}$ be the spectral data
associated with $q,r$. In particular,
the divisor $\mathcal{D}_{\underline{\hat{\mu}}(0,0)},\mathcal{D}_{\underline{\hat{\nu}}(0,0)}$
are nonspecial. Moreover, suppose the symmetry
constraints
(\ref{311})-(\ref{313}) hold.
Then the solution $q$ given in (\ref{1227})
is the algebro-geometric solution of $FL_+$ hierarchy
if and only if
all the branch points $E_m<0,m=0,\ldots,2n+1,$
and
\begin{flalign}\label{3051a}
  \textrm{Re}(\textbf{Z})=-
  \frac{1}{2}Re \textbf{Q},\,\,(\textrm{mod}\,\, \mathbb{Z}^n).
\end{flalign}
 Under the constraint (\ref{3051a}), the initial value $q(0,0)$
is not arbitrary and should
be taken as the form
\begin{flalign}\label{3052a}
  q(0,0)=\frac{\theta(\textbf{Z}-\textbf{T})}{\theta(\textbf{Z})}e^{i\vartheta},
\end{flalign}
where $\vartheta\in\mathbb{R}$ is an arbitrary but fixed constant.
Finally, the algebro-geometric solutions of $FL_+$ hierarchy is given by
\begin{flalign}\label{3058}
q(x,t_{\underline{r}})=&\frac{\theta(\textbf{Z}-\textbf{T}
          +\textbf{V}x+\textbf{W}t_{\underline{r}})}{\theta(\textbf{Z}
          +\textbf{V}x+\textbf{W}t_{\underline{r}})}
          \times\exp\left(i(e_{0,-}-e_{0,+})x+i
       (\widetilde{\Omega}_{\underline{r}}^{0,-}-
       \widetilde{\Omega}_{\underline{r}}^{0,+})t_{\underline{r}}+i\vartheta\right).
\end{flalign}

\end{thm}
\proof
In ref. \cite{05} (cf. Theorem 5.1),
we have proved the
$j$-th components $V_j, W_j$ of vectors $\textbf{V}, \textbf{W}$
are explicitly written as
\begin{align}
 V_j=&-2i c_j(n) ,\label{3051}\\
 W_j =&2i(-1)^{n+1}\prod_{m=0}^{2n+1}E_m^{-1/2}
   \sum_{s=1}^{r_-}\tilde{c}_{r-s,-}\sum_{\ell=1}^{s}
   c_j(\ell)\hat{c}_{s-\ell}(\underline{E}^{-1})\nonumber\\
 &-2i\sum_{s=1}^{r_+}\tilde{c}_{r_+-s,+}
 \sum_{\ell=1}^{n}c_j(\ell)\hat{c}_{\ell+s-n-1}(\underline{E})\label{3052}
\end{align}
and hence one gets $\textbf{V}=-\overline{\textbf{V}}, \textbf{W}=-\overline{\textbf{W}}$.
Choose the integration paths $\gamma_{\pm}$
from $\widetilde{Q}_0$ to the point $P$
near near $P_{0,\pm}$, which do not intersect the cycles $a_j$ (see Figure 2)
and one finds
\begin{flalign}
\overline{\widetilde{\Omega}}_{\underline{r}}^{0,\pm}=&\lim_{P\rightarrow P_{0,\pm}}\Big(\int_{Q_0}^P\overline{\widetilde{\Omega}}_{\underline{r}}^{(2)}\mp
  \sum_{s=1}^{r_-}\tilde{c}_{r_--s}\zeta^{-2s}\Big)\nonumber\\
  =&\lim_{P\rightarrow P_{0,\pm}}\Big(\int_{\gamma_{\pm}}\tau^*\widetilde{\Omega}_{\underline{r}}^{(2)}\mp
  \sum_{s=1}^{r_-}\tilde{c}_{r_--s}\zeta^{-2s}\Big)\nonumber\\
  =&\lim_{P\rightarrow P_{0,\pm}}\Big(\int_{ \gamma_{\pm}}
  \widetilde{\Omega}_{\underline{r}}^{(2)}\mp
  \sum_{s=1}^{r_-}\tilde{c}_{r_--s}\zeta^{-2s}\Big)\nonumber\\
  =&\,\,~~\widetilde{\Omega}_{\underline{r}}^{0,\pm}.
\end{flalign}
Here we use $\gamma_{\pm}$
is invariant with respect to the action $\tau_1$, i.e. $\tau(\gamma_{\pm})=\gamma_{\pm}.$
The statement $\overline{e}_{0,\pm}=e_{0,\pm}$ is also true by using the similar procedure.
It should be also noticed, although we won't use it right now, that
$\underline{\Xi}_{Q_0}$ is half-period, that is,
\begin{equation}\label{a2.54}
 2\underline{\Xi}_{Q_0}=0,\quad \textrm{(mod $L_n$).}\end{equation}
Combining (\ref{z1}) with (\ref{a2.54})
then yields
\begin{flalign}
\overline{\underline{\Xi}}_{Q_0}=\underline{\Xi}_{Q_0}\in(\mathbb{Z}/2)^n.
\end{flalign}
According to the integration path between $P_{0,\pm}$ described in Figure 2,
the parameter $\textbf{T}$
satisfies
$\textbf{T}=\overline{\textbf{T}}$. Using above analysis about
$\textbf{V,W,T},\widetilde{\Omega}_{\underline{r}}^{0,\pm}, e_{0,\pm}$ and Proposition 1.1.2,
one finds
the condition $q(x,t_{\underline{r}})=\bar{r}(x,t_{\underline{r}})$
is equivalent to
\begin{flalign}\label{3118}
 q(0,0)&\frac{\theta(\textbf{Z})}
          {\theta(\textbf{Z}-\textbf{T})}\frac{\theta(\textbf{Z}-\textbf{T}
          +\textbf{V}x+\textbf{W}t_{\underline{r}})}{\theta(\textbf{Z}
          +\textbf{V}x+\textbf{W}t_{\underline{r}})} \nonumber\\
         &\times\exp\left(i(e_{0,-}-e_{0,+})x+i
       (\widetilde{\Omega}_{\underline{r}}^{0,-}-
       \widetilde{\Omega}_{\underline{r}}^{0,+})t_{\underline{r}}\right)\nonumber\\
 = \frac{1}{\overline{q(0,0)}}&\frac{\overline{\theta(\textbf{Z}-\textbf{T})}}
          {\overline{\theta(\textbf{Z})}}
          \frac{\overline{\theta(\textbf{Y}
          +\textbf{V}x+\textbf{W}t_{\underline{r}})}}
          {\overline{\theta(\textbf{Y}-\textbf{T}
          +\textbf{V}x+\textbf{W}t_{\underline{r}})}}\nonumber\\
         &\times\exp\left(i(e_{0,-}-e_{0,+})x+i
       (\widetilde{\Omega}_{\underline{r}}^{0,-}
       -\widetilde{\Omega}_{\underline{r}}^{0,+})t_{\underline{r}}\right)\nonumber\\
  = \frac{1}{\overline{q(0,0)}}&\frac{\theta(\overline{\textbf{Z}}-\textbf{T})}
          {\theta(\overline{\textbf{Z}})}
          \frac{\theta(-\overline{\textbf{Y}}
          +\textbf{V}x+\textbf{W}t_{\underline{r}})}{\theta(-\overline{\textbf{Y}}
          +\textbf{T}
          +\textbf{V}x+\textbf{W}t_{\underline{r}})}\nonumber\\
         &\times\exp\left(i(e_{0,-}-e_{0,+})x+i
       (\widetilde{\Omega}_{\underline{r}}^{0,-}
       -\widetilde{\Omega}_{\underline{r}}^{0,+})t_{\underline{r}}\right).
\end{flalign}
Then we get
\begin{flalign}\label{3065}
 \left|q(0,0)\frac{\theta{(\textbf{Z})}}{\theta(\textbf{Z}-\textbf{T})}\right|^2
 =\frac{\theta(\textbf{Z}
          +\textbf{V}x+\textbf{W}t_{\underline{r}})}{\theta(-\overline{\textbf{Y}}+\textbf{T}
          +\textbf{V}x+\textbf{W}t_{\underline{r}})}
          \frac{\theta(-\overline{\textbf{Y}}
          +\textbf{V}x+\textbf{W}t_{\underline{r}})}{\theta(\textbf{Z}-\textbf{T}
          +\textbf{V}x+\textbf{W}t_{\underline{r}})}.
\end{flalign}
The equality (\ref{3065}) is self-contained only under the
constraint
\begin{flalign}\label{3059a}
 \textbf{Z}=-\overline{\textbf{Y}}+\textbf{T}+\underline{m}+\underline{n}\Gamma,
\end{flalign}
for some $\underline{m},\underline{n}\in\mathbb{Z}^n$
and hence
(\ref{3065}) changes to
\begin{flalign}\label{3067}
 \left|q(0,0)\frac{\theta{(\textbf{Z})}}{\theta(\textbf{Z}-\textbf{T})}\right|^2
 =e^{-2\pi i<\underline{n},\textbf{T}>},
\end{flalign}
where we used the quasi-periodic property of Riemann
theta function:
$$\theta(\underline{z}+\underline{m}+\underline{n}\Gamma)=\theta(\underline{z})\exp\left(
-2\pi i<\underline{n},\underline{z}>-\pi i<\underline{n},\underline{n}\Gamma>
\right).$$
  The equality (\ref{3067}) makes sense if and only if $<\underline{n},\textbf{T}>=0\,\,(\textrm{mod}\,\,\mathbb{Z})$,
  which gives rise to (\ref{3052a}). Moreover, combing (\ref{1226})
  with (\ref{3059a}) yields
  \begin{equation}\label{3068a}
    \textbf{Z}=-\overline{\textbf{Z}}-\overline{\textbf{Q}}+\underline{m}+\underline{n}\Gamma.
  \end{equation}
 Thus,
 \begin{flalign}
   2 Re\textbf{Z}=-Re \textbf{Q}+\underline{m},\label{3069a}\\
   0=-Im \textbf{Q}+\underline{n}Im \Gamma.\label{3070a}
 \end{flalign}
 by taking real and imaginary
parts in (\ref{3068a}).
We claim that
\begin{equation}\label{3071}
  \underline{n}=0.
\end{equation}
To prove this conclusion, we first look at
the distribution of the branch points $E_m, m=0,\ldots,2n+1$
on the line.
By the condition $P_{0,\pm}=(0,\pm \frac{1}{4})\in\cur^1,$
one infers
the number of branch points in positive real axis
is even. Then using (\ref{a2.26}) and (\ref{a3053}) we find
the imaginary part of $\textbf{Q}$ is of the type $\frac{\underline{k}}{2}\Gamma, \underline{k}\in\mathbb{Z}^n$.
However, the solvability of (\ref{3070a})
requires $\underline{k}$=0,
i.e. $E_m<0, m=0,\ldots,2n+1,$
which indicates (\ref{3071}). Therefore, (\ref{3051a})
holds
by
relations (\ref{3069a})-(\ref{3071}).
Finally, expression
 (\ref{3058}) is the direct result of (\ref{1227}), (\ref{3052a}).
\qed\vspace{0.3cm}

\begin{rem}
The strategy in \textbf{Theorem \ref{lem3.3.1}}
is invalid when
 applied to
analyze
the reality condition for $q=-\overline{r}$.
In fact, in this case (\ref{3067}) changes to
\begin{flalign*}
  -\left|q(0,0)\frac{\theta{(\textbf{Z})}}{\theta(\textbf{Z}-\textbf{T})}\right|^2
 =e^{-2\pi i<\underline{n},\textbf{T}>}=1,
\end{flalign*}
which is meaningless.
\end{rem}

Next we consider the smoothness condition for the algebro-geometric solutions (\ref{3058})
of $FL_+$
hierarchy
constructed from (\ref{3051a}) and (\ref{3052a}).
\begin{thm}
Assume the conditions of Theorem \ref{lem3.3.1}
and the conclusions (\ref{3051a})-(\ref{3058}) hold.
Then the solution $q$ given in (\ref{3058})
is smooth if and only if $\textrm{Re}(\textbf{Q})\in(2\mathbb{Z})^n$.
\end{thm}
\proof The solution (\ref{3058}) can have the
singularities only on account of the zeros of
the Riemann theta function $\theta(\textbf{Z}+\textbf{V}x+\textbf{W}t_{\underline{r}})$
and
its argument $\textbf{Z}+\textbf{V}x+\textbf{W}t_{\underline{r}}$
belongs to the set $\{\chi\in J(\cur^1)|\tau_1^J \chi +\chi = \frac{1}{2}\textrm{diag}(H)\}$ (in this case $H$=0 ).
Here $\tau_1^J$ is the anti-holomorphic involution on Jacobian $J(\cur^1)$
induced by $\tau_1,$ given by ($\textrm{in this case} \,\, \tau_1(Q_0)=Q_0)$
\begin{flalign*}
  \tau_1^J(\chi)=\overline{\chi}+n_\chi \underline{\alpha}_{Q_0}(\tau_1(Q_0))),\,\,
\end{flalign*}
where $n_\chi\in\mathbb{Z}, n_\chi\leq n,$ is the degree of the divisor $\mathcal{D}$
such that $\underline{\alpha}_{Q_0}(\mathcal{D})=\chi.$ Then by \cite{01} (Corollary 4.3),
the solution (\ref{3058}) is smooth if
and only if $\textbf{Z}+\textbf{V}x+\textbf{W}t_{\underline{r}}\in i\mathbb{R}^n$
and hence $\textbf{Z}\in i\mathbb{R}^n$. We complete the proof.
\qed\vspace{0.3cm}

Next we shall consider the reality condition
for $q=-\overline{r}$.
In this case, we assume all
the branch points
$E_m, m=0,\ldots,2n+1$
of $\cur$
are pairwise
conjugate.
The antiholomorphic involution
is still defined by (\ref{313}) on $\cur$
and
the homology
basis is explicitly shown in Figure 3.
\begin{figure}[htbp]
       \begin{center}
       \includegraphics{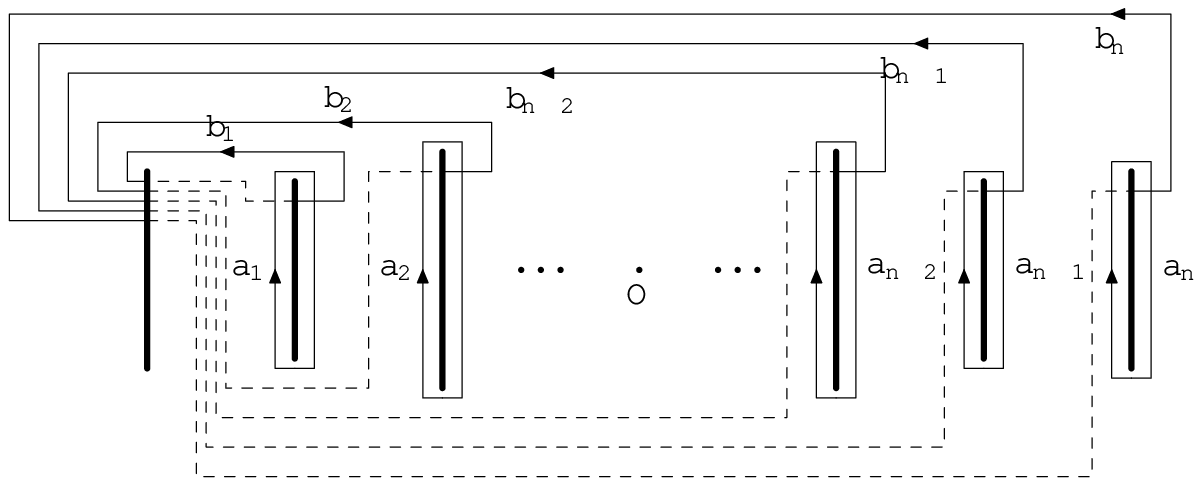}
      \caption{Homology basis on the real curves $\cur^2$, contours on sheet 1 are solid, contours on sheet 2
       are dashed}
      \end{center}
      \end{figure}

\begin{figure}[htbp]
       \begin{center}
       \includegraphics{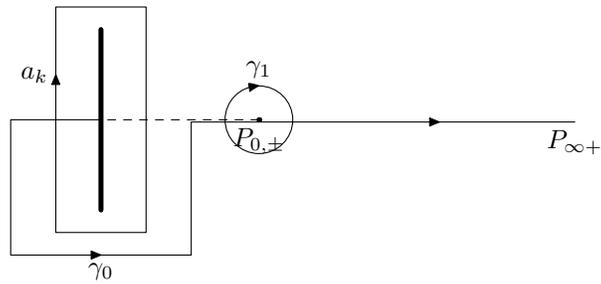}
      \caption{$\gamma_0$ is a path from $P_{0,-}$ to $P_{0,+}$ and $a_k$ is a
       cycle around the branch cut
       whose projection to $\xi$-sphere
       has negative real part and the minimal distance to 0}
      \end{center}
      \end{figure}

To avoid confusion we denote by $\cur^2$
the curve $\cur$ with pairwise distinct branch points.
It is easy to see that the holomorphic involution $\tau_1$
transforms the homology $\{a_j,b_j\}_{j=1}^n$ on $\cur^2$ as follows:
\begin{flalign}\label{3069}
 \tau_1(a_j)=-a_j,\,\,
 \tau_1(b_j)=b_j-\sum_{k\neq j}a_k.
\end{flalign}
From
\begin{flalign*}
&\int_{a_j}\overline{\eta}_k
=\int_{\tau_1(a_j)}\eta_k=\int_{a_j}\tau_1^*\eta_k,\,\,j,k=1,\ldots,n
\end{flalign*}
one gets $\tau_1^*\eta_k=\overline{\eta}_k,k=1,\ldots,n.$
In analogy to the process (\ref{3050}),
the conditions (\ref{112}) and (\ref{3069}) lead to
\begin{equation}\label{3070}
\tau^*\omega_j=-\overline{\omega}_j,\,\,j=1,\ldots,n.
\end{equation}
Then it follows $c_j(\ell)\in i\mathbb{R}, j,\ell=1,\ldots,n$ by (\ref{3.36}).
Using (\ref{3069}), (\ref{3070}), one finds
\begin{flalign}
 \overline{\int_{b_j}\omega_k}=&\int_{b_j}\overline{\omega}_k=-\int_{b_j}\tau_1^*\omega_k
 =-\int_{\tau_1(b_j)}\omega_k\nonumber\\
 =&-\int_{b_j}\omega_k+\sum_{
 \ell\neq j}\int_{a_\ell}\omega_k=-\int_{b_j}\omega_k+\sum_{\ell\neq j}\delta_{\ell k},
\end{flalign}
or equivalently,
\begin{flalign}
 \overline{\Gamma}=-\Gamma+\mathcal{P},\quad \mathcal{P}=
 (\mathcal{P}_{ij})_{n\times n},\quad \mathcal{P}_{ij}=1-\delta_{ij}.
\end{flalign}
By (\ref{112}),
(\ref{1226a}), (\ref{1227a}),
(\ref{20136901}), (\ref{20136902}) and (\ref{3069}),
one derives that
the constants $c_{j,\pm}^{(2s-1)},\tilde{c}_{j,\pm}^{(2s-1)}\in i\mathbb{R}$
and hence the action of $\tau_1$
on $\Omega_{P_{\infty\pm},2s-1}^{(2)},\Omega_{P_{0,\pm},2s-1}^{(2)}$
reads
\begin{flalign}
 \tau^*_1\Omega_{P_{\infty\pm},2s-1}^{(2)}=&\overline{\Omega}_{P_{\infty\pm},2s-1}^{(2)},\label{a2.24}\\
 \tau^*_1\Omega_{P_{0,\pm},2s-1}^{(2)}=&\overline{\Omega}_{P_{0,\pm},2s-1}^{(2)}.\label{a2.25}
 \end{flalign}
and hence
\begin{equation}\label{3075a}
  \tau_1^*\Omega_0^{(2)}= \overline{\Omega}_0^{(2)},\quad \tau_1^*\widetilde{\Omega}_{\underline{r}}^{0,\pm}=\overline{\widetilde{\Omega}}_{\underline{r}}^{0,\pm}.
\end{equation}

\begin{thm}\label{thm308}
Assume the algebro-geometric solution
$\{q,r\}\in\mathcal{M}^{\underline{n},\underline{r}}_0$ (see (\ref{1227}), (\ref{1228}))
and let
$\{\hat{\mu}_j(0,0), E_m;
q(0,0)\}$ be the spectral data
associated with $q,r$. In particular,
the divisor $\mathcal{D}_{\underline{\hat{\mu}}(0,0)},\mathcal{D}_{\underline{\hat{\nu}}(0,0)}$
are nonspecial. Moreover, suppose the symmetry
constraints
(\ref{311})-(\ref{313}) hold.
Then the expression $q$ given in (\ref{1227})
is the algebro-geometric solution of $FL_-$ hierarchy
if and only if
\begin{flalign}\label{3075}
 Im \textbf{Z}=-\frac{1}{2}Im \textbf{Q}+\underline{m} Im \Gamma,\,\,
    \underline{m}\in\mathbb{Z}^n\backslash (2\mathbb{Z})^n.
\end{flalign}
and in this constraint, the initial value $q(0,0)$
is not arbitrary and should
be taken as the form
\begin{flalign}\label{3076}
  q(0,0)=\frac{\theta(\textbf{Z}-\textbf{T})}{\theta(\textbf{Z})}\exp\left(-\pi <\underline{m},\textrm{Im}\textbf{T}>+i\varphi\right),
\end{flalign}
where $\varphi\in\mathbb{R}$ is an arbitrary but fixed constant.
Finally, the algebro-geometric solutions of $FL_-$ hierarchy is given by
\begin{flalign}\label{3077}
q(x,t_{\underline{r}})=&
\exp\left(-\pi <\underline{m},\textrm{Im}\textbf{T}>\right)
\frac{\theta(\textbf{Z}-\textbf{T}
          +\textbf{V}x+\textbf{W}t_{\underline{r}})}{\theta(\textbf{Z}
          +\textbf{V}x+\textbf{W}t_{\underline{r}})}\nonumber\\
          &\times\exp\left(i(e_{0,-}-e_{0,+})x+i
       (\widetilde{\Omega}_{\underline{r}}^{0,-}-
       \widetilde{\Omega}_{\underline{r}}^{0,+})t_{\underline{r}}+i\varphi\right).
\end{flalign}
\proof
 For the components of $\textbf{V}, \textbf{W}$ we get
\begin{flalign}
  \overline{V}_j=&-\frac{1}{2 \pi }\int_{b_j}\overline{\Omega}_0^{(2)}=-\frac{1}{2 \pi }\int_{b_j}\tau_1^*\Omega_0^{(2)}\nonumber\\
  =&-\frac{1}{2 \pi }\int_{\tau_1^*(b_j)}\Omega_0^{(2)}=-\frac{1}{2 \pi }\int_{b_j}\Omega_0^{(2)}\nonumber\\
  =&V_j,
\end{flalign}
and similarly $\overline{W}_j=W_j, j=1,\ldots,n.$ To determine value of the constant
$\textbf{T}$ let us
choose a specific path $\gamma_0$ from $P_{0,-}$ to $P_{0,+}$
which is shown in Figure 4. It is easy to see that
the holomorphic involution $\tau_1$ acts on the contour
$\gamma_0$ as follows,
\begin{equation}\label{3080}
  \tau_1(\gamma_0)=\gamma_0+a_{k}+\gamma_1,
\end{equation}
where $\gamma_1$ denotes a positively oriented circle around the point $P_{0,+}$ (see Figure 4).
It is clear that
\begin{flalign}
\overline{\widetilde{\Omega}}_{\underline{r}}^{0,\pm}=&\lim_{P\rightarrow P_{0,\pm}}\Big(\int_{Q_0}^P\overline{\widetilde{\Omega}}_{\underline{r}}^{(2)}\mp
  \sum_{s=1}^{r_-}\tilde{c}_{r_--s}\zeta^{-s}\Big)\nonumber\\
  =&\lim_{P\rightarrow P_{0,\pm}}
  \Big(\pm\frac{1}{2}\int_{\gamma_{0}}\tau^*\widetilde{\Omega}_{\underline{r}}^{(2)}\mp
  \sum_{s=1}^{r_-}\tilde{c}_{r_--s}\zeta^{-s}\Big)\nonumber\\
  =&\lim_{P\rightarrow P_{0,\pm}}\Big(\pm\frac{1}{2}\int_{\gamma_{0}+a_k+\gamma_1}
  \widetilde{\Omega}_{\underline{r}}^{(2)}\mp
  \sum_{s=1}^{r_-}\tilde{c}_{r_--s}\zeta^{-s}\Big)\nonumber\\
  =&\lim_{P\rightarrow P_{0,\pm}}\Big(\int_{Q_0}^P \widetilde{\Omega}_{\underline{r}}^{(2)}\mp
  \sum_{s=1}^{r_-}\tilde{c}_{r_--s}\zeta^{-s}\Big)\nonumber\\
  =&\,\,~~\widetilde{\Omega}_{\underline{r}}^{0,\pm}
\end{flalign}
and similarly we obtain $\overline{e}_{0,\pm}=e_{0,\pm}.$
From (\ref{3070}) and (\ref{3080}) one infers that
\begin{flalign}\label{3081}
 \overline{\textbf{T}}=&\overline{\underline{A}_{P_{0,-}}(P_{0,+})}=
 -\left(\int_{\gamma_0}\tau_1^*\omega_1,\ldots,\int_{\gamma_0}\tau_1^*\omega_n\right)\nonumber\\
 =&-\left(\int_{\tau_1^*(\gamma_0)} \omega_1,\ldots,\int_{\tau_1^*(\gamma_0)} \omega_n\right)\nonumber\\
 =&-\textbf{T}+I_{k},
  \end{flalign}
where $I_{k}=(0,\ldots,1\ldots,0)$ is the $k$-th
row of unit matrix $I.$
Next we try to search for the reality condition
for $q(x,t_{\underline{r}})=-\overline{{r}(x,t_{\underline{r}})}$.
Using Proposition \ref{prop113}, one derives
\begin{flalign}\label{3118}
 q(0,0)&\frac{\theta(\textbf{Z})}
          {\theta(\textbf{Z}-\textbf{T})}\frac{\theta(\textbf{Z}-\textbf{T}
          +\textbf{V}x+\textbf{W}t_{\underline{r}})}{\theta(\textbf{Z}
          +\textbf{V}x+\textbf{W}t_{\underline{r}})} \nonumber\\
         &\times\exp\left(i(e_{0,-}-e_{0,+})x+i
       (\widetilde{\Omega}_{\underline{r}}^{0,-}-
       \widetilde{\Omega}_{\underline{r}}^{0,+})t_{\underline{r}}\right)\nonumber\\
 = -\frac{1}{\overline{q(0,0)}}&\frac{\overline{\theta(\textbf{Z}-\textbf{T})}}
          {\overline{\theta(\textbf{Z})}}
          \frac{\overline{\theta(\textbf{Z}+\textbf{T}+\textbf{Q}
          +\textbf{V}x+\textbf{W}t_{\underline{r}})}}
          {\overline{\theta(\textbf{Z}+\textbf{Q}
          +\textbf{V}x+\textbf{W}t_{\underline{r}})}}\nonumber\\
         &\times\exp\left(i(e_{0,-}-e_{0,+})x+i
       (\widetilde{\Omega}_{\underline{r}}^{0,-}
       -\widetilde{\Omega}_{\underline{r}}^{0,+})t_{\underline{r}}\right)\nonumber\\
  = -\frac{1}{\overline{q(0,0)}}&
  \frac{\theta(\overline{\textbf{Z}}-\overline{\textbf{T}})}
          {\theta(\overline{\textbf{Z}})}\frac{\theta(\overline{\textbf{Z}}+\overline{\textbf{T}}
          +\overline{\textbf{Q}}
          +\textbf{V}x+\textbf{W}t_{\underline{r}})}{\theta(\overline{\textbf{Z}}+\overline{\textbf{Q}}
          +\textbf{V}x+\textbf{W}t_{\underline{r}})}\nonumber\\
         &\times\exp\left(i(e_{0,-}-e_{0,+})x+i
       (\widetilde{\Omega}_{\underline{r}}^{0,-}
       -\widetilde{\Omega}_{\underline{r}}^{0,+})t_{\underline{r}}\right),
\end{flalign}
i.e.
\begin{flalign}\label{3082}
 \left|q(0,0)\frac{\theta{(\textbf{Z})}}{\theta(\textbf{Z}-\textbf{T})}\right|^2
 =-\frac{\theta(\textbf{Z}
          +\textbf{V}x+\textbf{W}t_{\underline{r}})}{\theta(\overline{\textbf{Z}}
          +\overline{\textbf{Q}}
          +\textbf{V}x+\textbf{W}t_{\underline{r}})}
          \frac{\theta(\overline{\textbf{Z}}+\overline{\textbf{T}}+\overline{\textbf{Q}}
          +\textbf{V}x+\textbf{W}t_{\underline{r}})}{\theta(\textbf{Z}-\textbf{T}
          +\textbf{V}x+\textbf{W}t_{\underline{r}})}.
\end{flalign}
Taking into account (\ref{3081}), we conclude that
 \begin{flalign}\label{3083}
 \left|q(0,0)\frac{\theta{(\textbf{Z})}}{\theta(\textbf{Z}-\textbf{T})}\right|^2
 =-\frac{\theta(\textbf{Z}
          +\textbf{V}x+\textbf{W}t_{\underline{r}})}{\theta(\overline{\textbf{Z}}+\overline{\textbf{Q}}
          +\textbf{V}x+\textbf{W}t_{\underline{r}})}
          \frac{\theta(\overline{\textbf{Z}}-\textbf{T}+\overline{\textbf{Q}}
          +\textbf{V}x+\textbf{W}t_{\underline{r}})}{\theta(\textbf{Z}-\textbf{T}
          +\textbf{V}x+\textbf{W}t_{\underline{r}})}.
\end{flalign}
The equality (\ref{3083}) is self-contained only under the condition
\begin{equation}\label{3085a}
  \overline{\textbf{Z}}+\overline{\textbf{Q}}=\textbf{Z}+\underline{n}+\underline{m}\Gamma,\,\,
  \underline{n},\underline{m}\in\mathbb{Z}^n.
\end{equation}
By comparing the real part and imaginary part
of (\ref{3085a}) we arrive to
the relations
\begin{flalign}
&-2 Im \textbf{Z}=Im \textbf{Q}+\underline{m} Im \Gamma,\label{3086b}\\
&  \underline{n}+\frac{1}{2}\underline{m}
  \mathcal{P}=0.\label{3086a}
\end{flalign}
On the other hand,
from (\ref{3081}), (\ref{3083}) and (\ref{3085a}) one obtains
\begin{flalign}\label{3086}
 \left|q(0,0)\frac{\theta{(\textbf{Z})}}{\theta(\textbf{Z}-\textbf{T})}\right|^2
 =&-\exp\left(2\pi i<\underline{m},\textbf{T}>\right)\nonumber\\
 =&-\exp\left(2\pi i<\underline{m},\textrm{Re}\textbf{T}+i\textrm{Im}\textbf{T}>\right)\nonumber\\
 =&-\exp\left(\pi i<\underline{m},I_{k}\right)\exp\left(-2\pi <\underline{m},\textrm{Im}\textbf{T}>\right).
\end{flalign}
It is easy to see that
$<\underline{m}, I_{k}>=m_{k}$
and hence we know
the $k$-th component of $\underline{m}$
must be odd by comparing both sides of (\ref{3086}).
Taking into account (\ref{3086a}) one infers
\begin{equation}\label{3088}
 \underline{m}\in\mathbb{Z}^n\backslash (2\mathbb{Z})^n.
\end{equation}
Moreover, by (\ref{3086}) and (\ref{3088}) we get
\begin{flalign}\label{3087}
 \left|q(0,0)\frac{\theta{(\textbf{Z})}}{\theta(\textbf{Z}-\textbf{T})}\right|^2
 =\exp\left(-2\pi <\underline{m},\textrm{Im}\textbf{T}>\right).
\end{flalign}
which indicates (\ref{3076}). (\ref{3077}) is the direct result of (\ref{1227}), (\ref{3076}).
\qed
\end{thm}

\subsection*{Acknowledgment}
The work described in this paper
was supported by grants from the National Science
Foundation of China (Project No.11271079), Doctoral Programs Foundation of
the Ministry of Education of China, and the Shanghai Shuguang Tracking Project (project 08GG01).


\bibliographystyle{amsplain}

\end{document}